# New Fast Transform for Orthogonal Frequency Division Multiplexing

**Said Boussakta[a], Mounir T. Hamood[b] and Mohammed Sh. Ahmed[c]**

[a] School of Engineering, Newcastle University, Newcastle upon Tyne, NE1 7RU, UK
said.boussakta@newcastle.ac.uk
[b] Electrical Engineering Department, Tikrit University, Tikrit, P. O. BOX 42, IRAQ
m.t.hamood@tu.edu.iq
[c] Control of Petroleum Systems Engineering Department, Tikrit University, Tikrit, P. O. BOX 42, IRAQ
mohammed.shwash@tu.edu.iq

**Abstract**

In this paper, a new fast and low complexity transform is introduced for orthogonal frequency division multiplexing (OFDM) wireless systems. The new transform combines the effects of fast complex-Walsh-Hadamard transform (CHT) and the fast Fourier transform (FFT) into a single unitary transform named in this paper as the complex transition transform (CTT). The development of a new algorithm for fast calculation of the CT transform called FCT is found to have all the desirable properties such as in-place computation, simple indexing scheme and considerably lower arithmetic complexity than existing algorithms. Furthermore, a new OFDM system using the FCT algorithm is introduced and its performance has been evaluated. The proposed CT-OFDM achieves a noticeable reduction in peak-to-average-power-ratio (PAPR) and a significant improvement in the bit-error-rate (BER) performance compared with the conventional OFDM.

## 1. Introduction

Over the years, fast transforms have played a major role in wireless communications and signal processing. Recently, there has been intensive research to use combined transforms in an attempt to reduce the PAPR and BER in OFDM [1-4]. In [5] the authors showed that an OFDM precoded-WHT has many advantages and can achieve full frequency diversity gain. However, these techniques are still cumbersome involving the calculation of two transforms Walsh-Hadamard transform followed by the IFFT which leads to higher computation time and complexity. Single transform techniques are found to be more advantageous in terms of improving the PAPR and BER performance whilst also reducing the computation complexity [6-8].

Therefore, in this letter, a new low complexity single unitary transform is introduced. This transform is called the complex-transition transform (CTT). The CTT has the same effects of the CHT and FFT combined, but it has a much simpler structure and faster calculation. In particular, the new transform is found to be more efficient than conventional transforms in OFDM systems offering three advantages; lower complexity, lower PAPR and BER. This is because the new transform has an inherent ability of spreading all the individual information symbols over all the subcarriers, so highly attenuated subcarriers can be recovered from other subcarriers, resulting in an increase in the reliability of communications.

The paper is organized as follows: Section 2 presents the definition of the CT and its relationship with DFT and CHT, followed by the complete development of new FCT algorithm in section 3. Section 4 assesses the algorithm's efficiency in terms of arithmetic operations, while section 5 evaluates the performance analysis of the proposed CT-OFDM system. Finally, a conclusion is given in section 6.

## 2. Complex Transition Transform (CTT)

The CTT combines the discrete Fourier transform (DFT) and the complex Walsh-Hadamard transform (CHT) together into a single unitary transform that has very efficient implementation. The CTT can be defined by forming a single matrix $CT[N]$ of order $N$ that joins the matrices of DFT and CHT together as follows

$$CT[N] = \frac{1}{N} DFT[N] \; CHT[N] \qquad (1)$$

where $DFT[N]$ is the discrete Fourier transform matrix, $CHT[N]$ is the complex Walsh-Hadamard matrix as defined in [17-19] and $N$ is power of two transform length. For clarity, we will briefly review CHT transform in next section.

## 2.1 Review of Complex Walsh-Hadamard Transform

The CHT $X(k)$ of $N$-periodic sampled data $x(n)$ is defined as:

$$X(k) = x(n) H_c(kn) \qquad k, n = 0, 1, ..., N-1 \tag{2}$$

where $H_c(kn)$ is an $N \times N$ unitary matrix whose elements are defined by:

$$H_c(kn) = (-1)^{kn}(-j)^{k\dot{n}/2}$$
$$kn = \sum_{m=0}^{l-1}(k_m \cdot n_m) \tag{3}$$

In (3), the term $(k_m \cdot n_m)$ denotes the bit-by-bit module 2 product of the integers $k$ and $n$ respectively, $l=\log_2 N$ is the number of binary digit in each index, $\dot{n}/2$ is the binary representation of the highest power of two in $n/2$. Therefore, the CHT matrix of order $N$ can be generated recursively for $H_c[1]=1$, as follows:

$$H_c[N] = \begin{bmatrix} H_c[N/2] & H_c[N/2] \\ L \otimes H[N/4] & -L \otimes H[N/4] \end{bmatrix} \qquad N \geq 4 \tag{4}$$

$$\text{where } L = \begin{bmatrix} 1 & -j \\ 1 & j \end{bmatrix} \text{ and } H_c[2] = \begin{bmatrix} 1 & 1 \\ 1 & -1 \end{bmatrix}$$

The operator $\otimes$ in (4) implies the Kronecker product and $H(N)$ denotes the WHT matrix defined as:

$$H[N] = \begin{bmatrix} H[N/2] & H[N/2] \\ H[N/2] & -H[N/2] \end{bmatrix} \qquad N \geq 2 \tag{5}$$

Fast algorithms can be developed for the CHT transform, they include either matrix partitioning or factoring techniques [17, 20], that result in efficient algorithms for fast computation of CHT (FCHT). Fig. 1 shows the signal flow graph (SFG) of the 8 point FCHT.

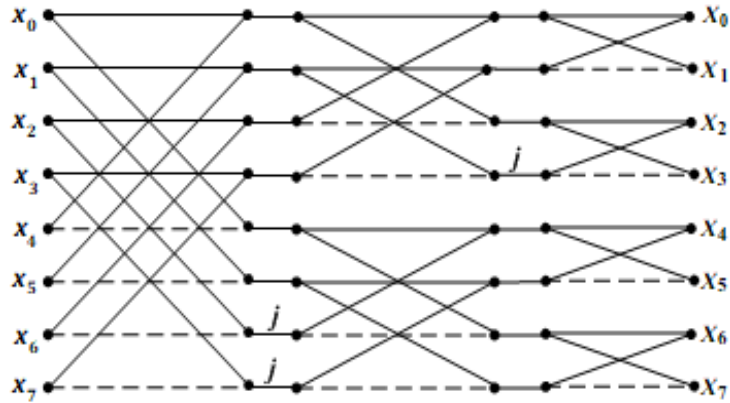

Fig. 1. Signal flow graph for 8-point radix-2 FCHT

## 3. Development of FCT Algorithm

Direct computation of the combination of DFT and CHT transforms needs $N^2$ complex multiplications and $2N(N-1)$ complex additions, which are considerably high and hence fast complex Hadamard-Fourier transform (FCHFT) need to be developed. As an example and without loss of generality, let us use an input sequence of length $N=8$ to explains this algorithm clearly. The SFG of the FCHT is as shown in Fig. 1, and the SFG for radix-2 FFT algorithm is as shown in Fig.2.Therefore, the resultant joint graph for the combined radix-2 FFT and CHT algorithms is as shown in Fig. 3. It should be noted that the solid and dotted lines in these figures stand for additions and subtractions respectively. Although the arithmetic complexity using fast algorithms based on cascading FCHT and FFT transforms is reduced by order of $\log_2 N$ complex operations, they are still high and need to be reduced for this method to be practical. Therefore, fast Complex T-transform (FCT) transform algorithm is developed in next section.

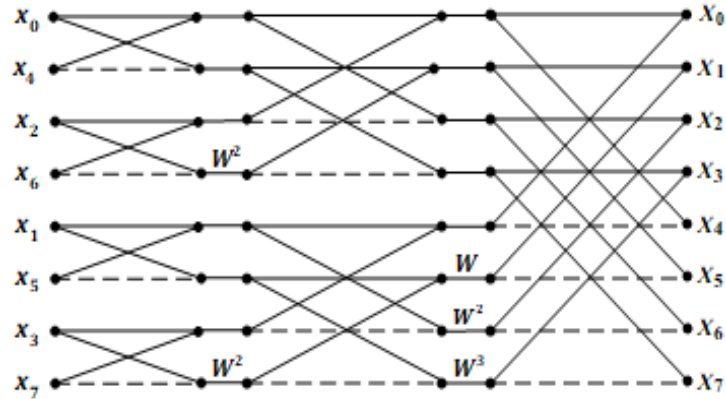

Fig. 2. Signal flow graph for 8-point radix-2 FFT

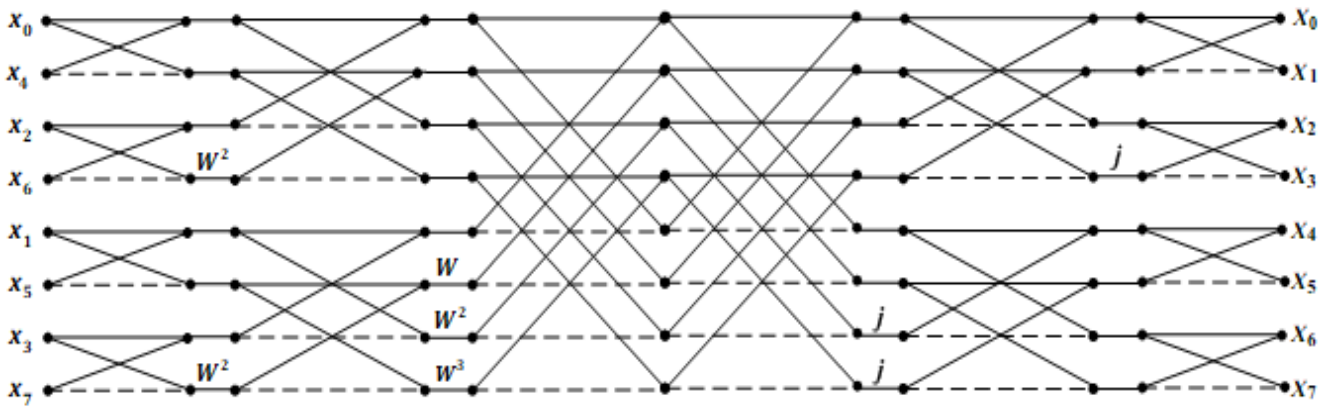

Fig. 3. Signal flow graph for the 8-point radix-2 FFT algorithm joint with the CHT

## 3.1 Fast Complex T-transform (FCT) Algorithm

The CHT matrix given in (4) can be written in a more convenient form as

$$H_c[N] = \begin{bmatrix} H_c[\frac{N}{2}] & S[\frac{N}{2}]H[\frac{N}{2}] \\ H_c[\frac{N}{2}] & -S[\frac{N}{2}]H[\frac{N}{2}] \end{bmatrix},$$

$$S[\frac{N}{2}] = \begin{bmatrix} I[\frac{N}{4}] & 0 \\ 0 & jI[\frac{N}{4}] \end{bmatrix}$$

(6)

Also the DFT matrix in row bit reverse order $\hat{F}[N]$ can be written in terms of lower matrices of order ($N/2$) as

$$\hat{F}[N] = \begin{bmatrix} \hat{F}[\frac{N}{2}] & \hat{F}[\frac{N}{2}] \\ \hat{F}[\frac{N}{2}]D[\frac{N}{2}] & -\hat{F}[\frac{N}{2}]D[\frac{N}{2}] \end{bmatrix}$$

(7)

where $D\left[\frac{N}{2}\right]$ is a diagonal matrix whose elements have values of $W^i = exp\left(-j\frac{2\pi i}{N}\right)$ for $i = 0, 1, \ldots, \frac{N}{2} - 1$.

Substituting (6) and (7) into (1), gives

$$CT[N] = \frac{1}{N} \begin{bmatrix} \hat{F}[\frac{N}{2}] & \hat{F}[\frac{N}{2}] \\ \hat{F}[\frac{N}{2}]D[\frac{N}{2}] & -\hat{F}[\frac{N}{2}]D[\frac{N}{2}] \end{bmatrix} \begin{bmatrix} H_c[\frac{N}{2}] & S[\frac{N}{2}]H[\frac{N}{2}] \\ H_c[\frac{N}{2}] & -S[\frac{N}{2}]H[\frac{N}{2}] \end{bmatrix}$$

$$= \frac{2}{N} \begin{bmatrix} \hat{F}[\frac{N}{2}]H_c[\frac{N}{2}] & 0 \\ 0 & \hat{F}[\frac{N}{2}]D[\frac{N}{2}]S[\frac{N}{2}]H[\frac{N}{2}] \end{bmatrix}$$

(8)

Equation (8) shows that the CT matrix of order $N$ can be written in terms of lower order matrices as

$$CT[N] = \begin{bmatrix} CT[\frac{N}{2}] & 0 \\ 0 & \frac{2}{N}\left[\hat{F}[\frac{N}{2}]D[\frac{N}{2}]S[\frac{N}{2}]H[\frac{N}{2}]\right] \end{bmatrix}$$

(9)

The lower part of (9) can be factorized further and the $CT[N/2]$ is the matrix of CT transform of order $N/2$ and can be factorized further. Therefore, the CT matrix can be transformed into a product of sparse matrices leading to a fast calculation of CT transform (FCT). For example and without loss of generality, for a transform length of $N=8$, Eq. (8) can be written as

$$CT[8] = \frac{1}{4} \begin{bmatrix} \hat{F}[4]H_c[4] & 0 \\ 0 & \hat{F}[4]D[4]S[4]H[4] \end{bmatrix}$$

(10)

For DFT matrix of length $N=8$, $\hat{F}[4]$ in (10) can be written as

$$\hat{F}[4] = \begin{bmatrix} \hat{F}[2] & \hat{F}[2] \\ \hat{F}[2]D[2] & -\hat{F}[2]D[2] \end{bmatrix}$$

(11)

where $\hat{F}[2] = \begin{bmatrix} 1 & 1 \\ 1 & -1 \end{bmatrix}$ and $D[2] = \begin{bmatrix} 1 & 0 \\ 0 & w \end{bmatrix}$.

Using the fact $\left[W^{\frac{N}{4}} = -j\right]$, follows $\hat{F}[2]D[2] = \begin{bmatrix} 1 & -j \\ 1 & j \end{bmatrix}$, therefore (11) can be written as

$$\hat{F}[4] = \begin{bmatrix} 1 & 1 & 1 & 1 \\ 1 & -1 & 1 & -1 \\ 1 & -j & -1 & j \\ 1 & j & -1 & -j \end{bmatrix} \tag{12}$$

Similarly, the term $\hat{F}[4]D[4]$ in (10) can written as

$$\hat{F}[4]D[4] = \begin{bmatrix} 1 & 1 & 1 & 1 \\ 1 & -1 & 1 & -1 \\ 1 & -j & -1 & j \\ 1 & j & -1 & -j \end{bmatrix} \begin{bmatrix} 1 & 0 & 0 & 0 \\ 0 & w & 0 & 0 \\ 0 & 0 & -j & 0 \\ 0 & 0 & 0 & w^3 \end{bmatrix}$$

$$= \begin{bmatrix} 1 & w & -j & w^3 \\ 1 & -w & -j & -w^3 \\ 1 & w^3 & j & w \\ 1 & -w^3 & j & -w \end{bmatrix} \tag{13}$$

The upper half of (10) can be factorized further using (8) as

$$\frac{1}{4}\hat{F}[4]H_c[4] = \frac{1}{2}\begin{bmatrix} \hat{F}[2]H_c[2] & 0 \\ 0 & \hat{F}[2]D[2]S[2]H[2] \end{bmatrix} \tag{14}$$

Applying values of $\hat{F}[2], D[2], S[2], H[2]$ and that fact that $H_c[2]=H[2]$, Eq. (14) becomes

$$\frac{1}{4}\hat{F}[4]H_c[4] = \frac{1}{2}\begin{bmatrix} \begin{bmatrix} 1 & 1 \\ 1 & -1 \end{bmatrix}\begin{bmatrix} 1 & 1 \\ 1 & -1 \end{bmatrix} & 0 \\ 0 & \begin{bmatrix} 1 & 1 \\ 1 & -1 \end{bmatrix}\begin{bmatrix} 1 & 0 \\ 0 & -j \end{bmatrix}\begin{bmatrix} 1 & 0 \\ 0 & j \end{bmatrix}\begin{bmatrix} 1 & 1 \\ 1 & -1 \end{bmatrix} \end{bmatrix}$$

$$= \begin{bmatrix} 1 & 0 & 0 & 0 \\ 0 & 1 & 0 & 0 \\ 0 & 0 & 1 & 0 \\ 0 & 0 & 0 & 1 \end{bmatrix} = I[4] \tag{15}$$

where $I[4]$ is the identity matrix of order-4.

Using (13), the lower half of (10) can be simplified to

$$\frac{1}{4}\hat{F}[4]D[4]S[4]H[4] = \frac{1}{4}\begin{bmatrix} 1 & w & -j & w^3 \\ 1 & -w & -j & -w^3 \\ 1 & w^3 & j & w \\ 1 & -w^3 & j & -w \end{bmatrix}\begin{bmatrix} 1 & 0 & 0 & 0 \\ 0 & 1 & 0 & 0 \\ 0 & 0 & j & 0 \\ 0 & 0 & 0 & j \end{bmatrix}\begin{bmatrix} 1 & 1 & 1 & 1 \\ 1 & -1 & 1 & -1 \\ 1 & 1 & -1 & -1 \\ 1 & -1 & -1 & 1 \end{bmatrix}$$

$$= \begin{bmatrix} \frac{(1+w)}{2} & \frac{(1-w)}{2} & 0 & 0 \\ \frac{(1-w)}{2} & \frac{(1+w)}{2} & 0 & 0 \\ 0 & 0 & \frac{(1+w^3)}{2} & \frac{(1-w^3)}{2} \\ 0 & 0 & \frac{(1-w^3)}{2} & \frac{(1+w^3)}{2} \end{bmatrix}$$

(16)

Substituting (15) and (16) into (10), we obtain the fast CT matrix of order 8 as

$$CT[8] = \begin{bmatrix} I[4] & 0 \\ & \frac{(1+w)}{2} & \frac{(1-w)}{2} & 0 & 0 \\ 0 & \frac{(1-w)}{2} & \frac{(1+w)}{2} & 0 & 0 \\ & 0 & 0 & \frac{(1+w^3)}{2} & \frac{(1-w^3)}{2} \\ & 0 & 0 & \frac{(1-w^3)}{2} & \frac{(1+w^3)}{2} \end{bmatrix}$$

(17)

Therefore, the signal flow graph for 8-point radix-2 FCT algorithm is shown in Fig. 4.

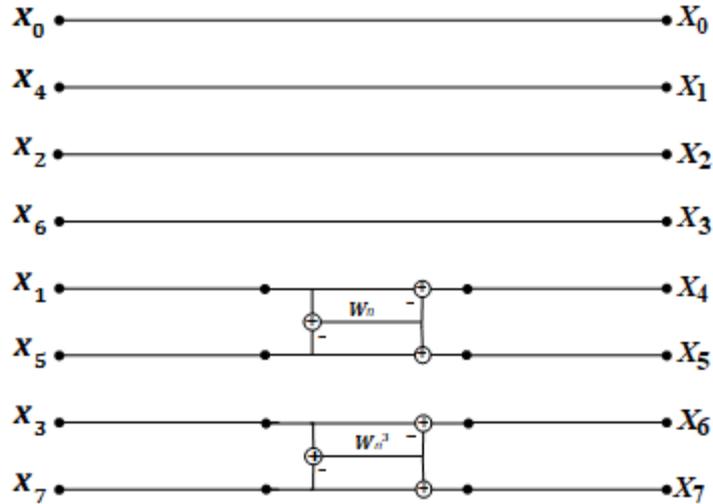

Fig. 4. Signal flow graph for the 8-point radiux-2 FCT algorithm where $W_n^i = (1-W^i)/2$

Based on the factorization in (17), we can construct the butterfly structure for this algorithm. Let $a$ and $b$ be the input of the CT transform's butterfly, the output $A$ and $B$ will be calculated as follows

$$\begin{aligned} A &= \frac{1}{2}\left[a(1+w^i)+b(1-w^i)\right] \\ &= a - \frac{1}{2}\left[(a-b)(1-w^i)\right] \end{aligned}$$

(18)

$$\begin{aligned} B &= \frac{1}{2}\left[a(1-w^i)+b(1+w^i)\right] \\ &= b + \frac{1}{2}\left[(a-b)(1-w^i)\right] \end{aligned}$$

The in-place butterfly that implements (18) is shown in Fig. 5, where ($i$) are arranged in bit-reversed order.

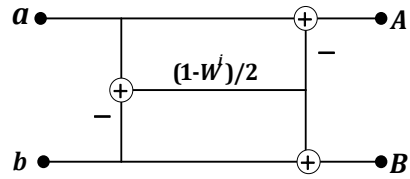

Fig. 5. In-place butterfly of the FCFT algorithm

Furthermore, the developed FCT algorithm is scalable and can be implemented for any power-of-two transform lengths. As an example, Fig. 6 shows the signal flow graph for the 16-point FCT algorithm, where the structure of each butterfly is exactly same as that depicted in Fig. 5.

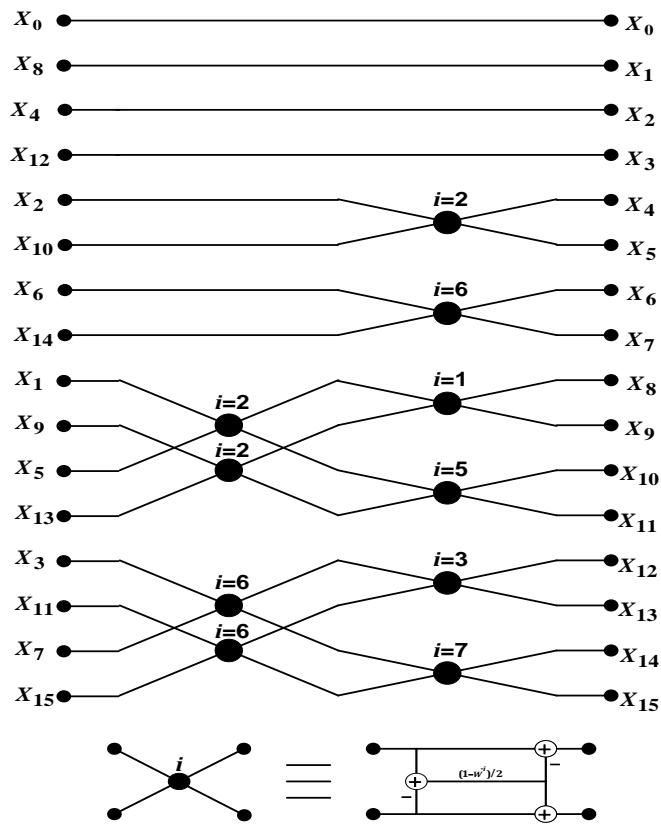

Fig. 6. Signal flow graph for the 16-point radiux-2 FCT algorithm

## 4. Arithmetic Complexity

As shown in Fig. 4, for $N=8$, the signal flow graph of the developed FCT algorithm needs 2 complex multiplications and 6 complex additions, while the algorithm based on cascading CHT followed by FFT (CHT-FFT) shown in Fig.3 needs 8 complex multiplications and 48 complex additions. Also, the calculation of CHT-FFT for $N=8$ points involves 6 stages, while the FCT only involves one stage. This leads to great savings in terms of arithmetic operations and hardware complexity. In general, each of the butterflies shown in Fig. 5 calculates two points and requires one multiplication and three additions. The whole transform needs ($\log_2 N$-2) stages and the total number of butterflies saved is equal to (3$N$/2-2). Therefore, the total number of complex multiplications $M_{c1}$ and complex additions $A_{c1}$ for the FCT algorithm can be calculated as

$$\begin{aligned} M_{c1} &= \tfrac{N}{2}\log_2 N - \tfrac{3N}{2} + 2 \\ A_{c1} &= \tfrac{3N}{2}\log_2 N - \tfrac{9N}{2} + 6 \end{aligned} \tag{19}$$

The corresponding number of complex multiplications $M_{c2}$ and complex additions $A_{c2}$ for fast calculation of CHT-FFT algorithm based on similar implementation (Radix-2), without counting trivial multiplications, are given as

$$\begin{aligned} M_{c2} &= \tfrac{N}{2}\log_2 N - \tfrac{N}{2} \\ A_{c2} &= 2N \log_2 N \end{aligned} \tag{20}$$

A comparison of (19) and (20), reveals that the proposed algorithm involves ($N$-2) fewer complex multiplications and [($N\log_2 N$+9$N$-12)/2] fewer complex additions. The number of arithmetic operations for the developed FCT and the fast CHT-FFT algorithms are as shown in Table 1 for different transform lengths. The results show a massive reduction in arithmetic operations and hardware complexity for the developed FCT algorithm.

TABLE I
ARITHMETIC OPERATIONS FOR FCFT AND CHT-FFT ALGORITHMS

| Length | FCT Algorithm | | CHT-FFT Algorithm | | Total saving |
|---|---|---|---|---|---|
| $N$ | $M_{c1}$ | $A_{c1}$ | $M_{c2}$ | $A_{c2}$ | % |
| 8 | 2 | 6 | 8 | 48 | 86 |
| 16 | 10 | 30 | 24 | 128 | 74 |
| 32 | 34 | 102 | 64 | 320 | 65 |
| 64 | 98 | 294 | 160 | 768 | 58 |
| 128 | 258 | 774 | 384 | 1792 | 53 |
| 256 | 642 | 1926 | 896 | 4096 | 48 |
| 512 | 1538 | 4614 | 2048 | 9216 | 45 |
| 1024 | 3586 | 10758 | 4608 | 20480 | 43 |

## 5. Performance of the CT-OFDM system

In this section, the application of the new CT-transform to OFDM is discussed. For simplicity, the results presented in this section are obtained based on assumptions of perfect knowledge of channel response, and perfect frequency and time synchronizations:

## 5.1 Peak-to-Average Power Ratio (PAPR)

In OFDM systems, the input signals are processed through ($\log_2 N$) stages of the IFFT. Therefore, owing to the superposition of signals, the peak power of the output signals can be high when compared to their average power. In the proposed CT-OFDM system, the inverse complex transition transform (ICTT) involves ($\log_2 N$-2) stages only and four direct paths. The maximum number of non-zero elements in each section of the ICTT is less than (*N*/4). Consequently, the superposition of the input signals processed through the new ICTT is less than in the case of the IFFT, leading to a lower PAPR compared with both conventional OFDM and T-OFDM [6] systems as shown in Fig. 7. Due to the unitary characteristics of the CT-transform, peak reduction is achieved with the preservation of the average power for the transmitted samples leading to much lower PAPR.

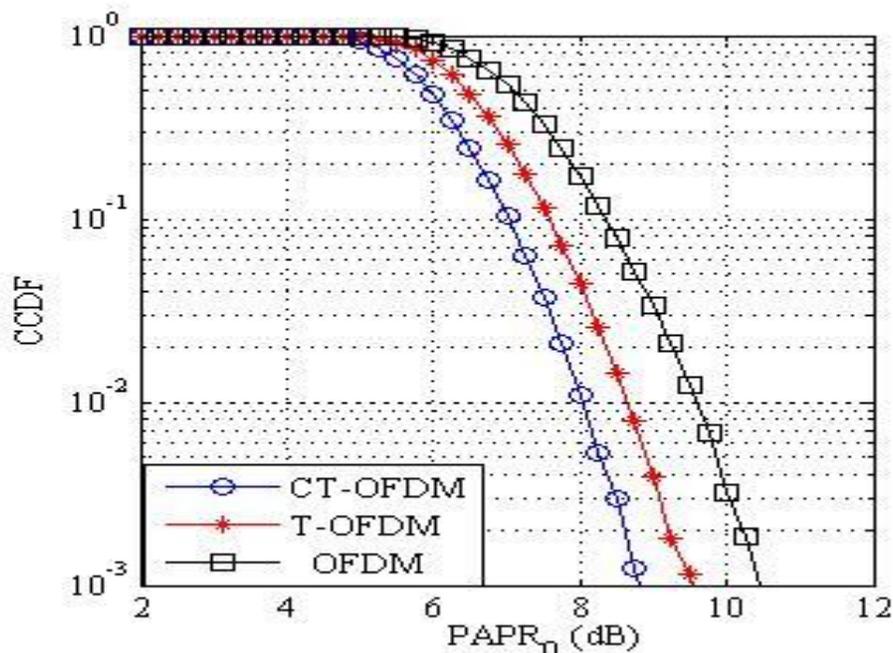

Fig. 7. CCDF for conventional OFDM, T-OFDM, and CT-OFDM systems.

## 5.2 Bit Error Rate (BER)

BER is a typical performance measure for quantifying the benefits of using the proposed CT-OFDM over the conventional OFDM system. In order to investigate the BER performance of such systems over multipath transmission, the following parameters are utilized: number of subcarriers is 1024, the cyclic prefix (CP) length is 256, the sample time is 88 ns and the system bandwidth is 10 MHz. It should be noted that the quasi-static 6-taps ITU-B (pedestrian) channel model [11, 12] is adopted to evaluate the BER performance of the proposed system in this paper. The BER performance of the investigated CT-OFDM system with QPSK and 16-QAM modulation is demonstrated in Fig. 8. It is clear from this figure that the CT-OFDM scheme exhibits significant improvement in the BER performance over the multipath propagation when compared with the conventional OFDM system. The improved performance is a result of the inherent ability of the CT transform to achieve frequency diversity by spreading each subcarrier over others. These high diversity systems mitigate the deep fade effect arising from the multipath channel on an individual subcarrier.

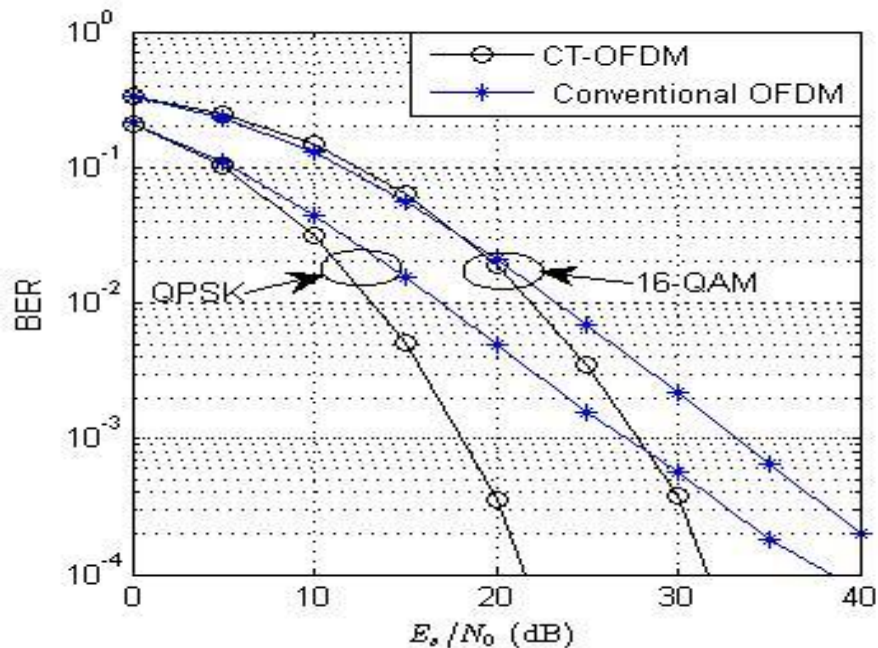

Fig. 8. BER performance of MMSE equalizer for OFDM systems across ITU-B channel with 16-QAM and QPSK mapping.

## 6. Conclusion

This paper proposed a new fast and efficient transform that combines the effects of the CHT and the FFT transforms into a single unitary transform named as the complex transition (CT) transform. Furthermore, a fast algorithm for the computation of CT-transform (FCT) has been developed showing that the new transform has a very low complexity and efficient implementation. The newly developed FCT algorithm outperforms CHT-FFT algorithm in terms of arithmetic operations and offers a significant complexity reduction. Moreover, the FCT algorithm can be implemented in-place using an integrated butterfly structure, and is suitable for both software and hardware implementations. The application of the developed transform to OFDM has been found to offer great potential in terms of BER performance improvement and PAPR reduction for both QAM and QPSK modulation techniques.